\documentclass{article}
\usepackage{amsmath}
\usepackage{graphicx}
\usepackage{natbib}
\usepackage{url} 
\usepackage{graphicx}
\usepackage{amsmath}
\usepackage{natbib}
\usepackage{amscd,amssymb,amsfonts,verbatim}
\usepackage{mathrsfs}
\usepackage{bbm}
\usepackage{listings}
\usepackage{epsfig}
\usepackage{enumitem}
\usepackage{subcaption}
\usepackage[plain]{algorithm2e}
\usepackage{algorithmic}
\usepackage{url}
\usepackage{booktabs,makecell,ltablex}
\usepackage{multirow}

\usepackage{tikz}

\usetikzlibrary{shapes,decorations,arrows,calc,arrows.meta,fit,positioning}

\usepackage{pgfplots}

\usepackage{xcolor}
\tikzset{
	-Latex,auto,node distance =1 cm and 1 cm,semithick,
	state/.style ={circle, draw, minimum width = 0.7 cm},
	box/.style ={rectangle, draw, minimum width = 0.7 cm, fill=lightgray},
	point/.style = {circle, draw, inner sep=0.08cm,fill,node contents={}},
	bidirected/.style={Latex-Latex,dashed},
	el/.style = {inner sep=3pt, align=left, sloped}
	
}

\def\E{\mathbb{E}}

\def\E{\mathbb{E}}

\renewcommand\hat{\widehat}

\def\E{\mathbb{E}}
\newcommand{\indep}{\perp \!\!\! \perp}

\def\b0{\boldsymbol{0}}

\graphicspath{{Figs/}}

\makeatother

\begin{document}

\def\spacingset#1{\renewcommand{\baselinestretch}%
{#1}\small\normalsize} \spacingset{1}


  \title{\bf A longitudinal Bayesian framework for estimating causal dose-response relationships}
  \author{Yu Luo\thanks{Department of Mathematics, King's College London, U.K.}, \hspace{.2cm}
Kuan Liu\thanks{Institute of Health Policy, Management and Evaluation, University of Toronto, Canada.}, \hspace{.2cm} Ramandeep Singh\thanks{Chair of Transportation Systems Engineering, Technical University of Munich, Germany.}, \hspace{.2cm} Daniel J. Graham\thanks{Department of Civil and Environmental Engineering, Imperial College London, U.K.} \hspace{.2cm} \\
  }
 \date{ }
  \maketitle

\bigskip
\begin{abstract}
Existing causal methods for time-varying exposure and time-varying confounding focus on estimating the average causal effect of a time-varying binary treatment on an end-of-study outcome, offering limited tools for characterizing marginal causal dose-response relationships under continuous exposures. We propose a scalable, nonparametric Bayesian framework for estimating marginal longitudinal causal dose-response functions with repeated outcome measurements. Our approach targets the average potential outcome at any fixed dose level and accommodates time-varying confounding through the generalized propensity score. The proposed approach embeds a Dirichlet process specification within a generalized estimating equations structure, capturing temporal correlation while making minimal assumptions about the functional form of the continuous exposure. We apply the proposed methods to monthly metro ridership and COVID-19 case data from major international cities, identifying causal relationships and the dose-response patterns between higher ridership and increased case counts.
\end{abstract}

\noindent%
\vfill

\newpage
\spacingset{1.5}

\section{Introduction}
The SARS-CoV-2 virus pathogen, which causes the COVID-19 pandemic, is considered to transmit via two main pathways: (i) directly through droplets and aerosols, and (ii) indirectly through fomite transmission \citep[for example,][]{Derqui2023}. Public transport modes serving dense urban areas induce prolonged exposure for passengers in enclosed and oftentimes poorly ventilated spaces, and they also facilitate the mixing of populations at the origin and destination trip-ends. Consequently, we can reasonably hypothesize that public transport is a vector of virus transmission in a city. While this hypothesis seems intuitively reasonable, and has been put forward previously, including in the mass media \citep{thomas2022investigating}, it remains largely unsubstantiated statistically under the causal lens. Existing empirical studies quantifying the dynamic impact of mass public transport on COVID-19 transmission are scarce. Most previous studies have analyzed aggregate human mobility, rather than mass transit ridership specifically, for instance, using less granular mobile phone data sources such as those available via Google Mobility or Apple Maps. Despite growing evidence of multi-layered causal pathways, the causal dynamics connecting mass transit ridership, such as metro ridership, to infectious-disease spread remain underexplored.

We consider a longitudinal causal inference setting indexed by time $t$, where $Y_t$ denotes the outcome (COVID-19 cases in the city), $D_t$ the treatment (urban mass transit ridership), $X_t$ a set of time-varying confounders, and $U$ unobserved time-invariant heterogeneity.  Our objective is to estimate the marginal causal effect of $D_t$ on $Y_t$. This presents two key challenges. First, we must appropriately adjust for observed confounding covariates $X_t$, while allowing for the potential influence of unobserved time-invariant heterogeneity $U$. These include time-varying (i.e., $X_t$) factors such as the severity of containment measures on human mobility (e.g. lockdown measures), urban mobility characteristics, and vaccination rates, along with a set of unmeasured time-invariant variables (i.e. $U$) city level attributes such as characteristics of the built-environment, urban culture, and baseline socio-demographics. Second, the data involve a dynamic longitudinal structure, with repeated outcome measurements, time-varying treatment exposures (doses), and a collection of dynamic confounders that change over time.

Most existing methods for causal estimation of the effect of a continuous-exposure are developed for the point-treatment framework, including the generalized propensity score (GPS)-based approaches \citep{robins2000marginal, hirano2004propensity,moodie2012estimation,galvao2015uniformly}, g-computation and outcome regression-based approaches \citep{hill2011bayesian}, and a doubly robust estimator considering combining both treatment and outcome models \citep{kennedy2017non}. Extensions to longitudinal causal data are very limited and restricted to the parametric frequentist paradigms \citep{robins2000marginal, vansteelandt2016revisiting}. In particular, g-computation can be computationally intensive and intractable when estimating treatment regime effects in longitudinal settings with high-dimensional time-varying confounding, as it requires fitting sequential models of the complete data-generating process and simulating counterfactual trajectories of time-varying covariates and outcomes at a set of treatment levels \citep{robins1986new, linero2022simulation}. On the other hand, the Bayesian lens for longitudinal causal dose-response analysis is largely missing and remains underexplored in the literature. The Bayesian estimation framework can offer both statistical and practical advantages, including the propagation of uncertainty through complex model structures and the ability to provide probabilistic summaries of causal effects, which facilitates decision-making in applied settings such as our ridership--COVID study. In addition, frequentist methods such as generalized estimating equations (GEEs) do not naturally accommodate prior information. These features motivate a Bayesian alternative that retains the scalability of estimating-equation methods while enabling posterior summaries of longitudinal dose--response effects.  The incorporation of prior beliefs under the proposed Bayesian approach can be particularly valuable in biomedical and policy applications where historical evidence or expert knowledge is available. Moreover, unlike approaches that rely on large-sample approximations, Bayesian inference provides coherent uncertainty quantification even in small-sample settings, which commonly arise in dose–response analyses. These considerations motivate us to develop Bayesian alternatives for longitudinal causal dose–response modeling. In this paper, therefore, we fill this literature gap by introducing a flexible yet computationally tractable non-parametric Bayesian estimation framework that estimates the entire longitudinal dose-response surface while making minimal distribution assumptions about the functional form of the continuous exposure.

Our specific contributions in this paper are on three fronts: i) we propose a nonparametric Bayesian framework using the Dirichlet process (DP) for marginal causal dose–response estimation with longitudinal data, which integrates GEEs with coherent Bayesian uncertainty quantification; ii) we demonstrate how the proposed DP–based framework accommodates multiple causal adjustment strategies, including covariate-adjusted outcome regression and inverse probability weighted marginal structural models; and iii) we provide computationally scalable algorithms and open-source code (\url{https://github.com/yumcgill/Bayes_DoseResponse}) making the proposed Bayesian approach practical for large, real-world datasets for wide adoption and application beyond the motivating study.

The remainder of this paper is organized as follows. In Section \ref{sec:meth}, we review methods for causal analysis of longitudinal data with repeated outcomes and introduce a non-parametric Bayesian formulation for inference when the model is specified through GEEs. Section \ref{sec:sim} presents simulation studies, comparing a weighting-based method with an approach that incorporates the GPS as a covariate. In Section \ref{sec:real}, we apply our methods to a real data example, analyzing the causal effect of mass transit ridership on the spread of COVID-19. Finally, we conclude with some remarks in Section \ref{sec:dis}.

\section{Methodology}
\label{sec:meth}
\subsection{Causal inference for marginal dose-response functions with repeated outcomes}
\label{sec:apo}
In the longitudinal causal inference setting, let $t=1,\ldots,T$ index time. Let $Y_t$ denote the outcome, $D_t\in\mathbb{R}$ denote the (possibly continuous) dose at time $t$, $X_t$ denote a set of time-varying observed covariates,  and $U$ denote unobserved time-invariant heterogeneity, where dependence on treatment assignment operates through the observed covariates $X_t$.  Therefore, the observed data consist of the longitudinal process $Z = \{(Y_t,X_t,D_t):t = 1,\ldots, T\}$. Using the potential outcome framework for longitudinal data, we define $Y_t(d_t)$ as the counterfactual outcome at time $t$ that would be observed if a study unit were exposed to the dose $d_t$ at time $t$. Our primary estimand is a  marginal dose-response function, defined as
\[
\mu(d) = \mathbb{E}\!\left\{ Y(d) \right\},
\]
where $Y(d)$ denotes the potential outcome that would be observed if, at each time point, the exposure were set to the constant level $d$.  This estimand represents the average potential outcome (APO) under an exposure level $d$. In other words, the estimand depends only on the marginal level of exposure $d$, rather than on the full longitudinal treatment path used to achieve it. This marginal dose-response parameterization has commonly been used in transportation engineering. As in the transportation setting, it is neither feasible nor substantively meaningful to define a single, well-specified longitudinal treatment regime. Instead, our scientific goal is to quantify the APO at a given ridership on COVID-related cases. We believe it is suitable to estimate the dose-response relationship in our COVID analysis. For example, it was employed by \cite{graham2014quantifying} in their analysis to quantify the effect of the network capacity expansions on urban traffic volumes. Therefore, we adopt this estimand to characterize the dose-response relationship in our COVID analysis.

To identify the APO, $\mu(d)$, we invoke the following assumptions. First, we assume no interference across units, i.e., for each study unit $i$ ($i=1,\ldots,n$), $Y_i(d_1,\ldots,d_n)=Y_i(d_i)$, corresponding to the stable unit treatment value assumption (SUTVA). Second, we assume consistency, that is, if an study unit is observed to receive dose $D_t=d$ at time $t$, then the observed outcome satisfies $Y_t=Y_t(d)$. Third, we assume marginal ignorability of treatment assignment, that is, at each time $t=1,\ldots,T$, $Y(d) \;\perp\!\!\!\perp\; D_t \mid X_t$. This assumption does not condition on past dose, and allows past dose to affect both $D$ and $Y$ indirectly, provided those effects are captured by $X$. Fourth, we assume positivity. For any covariate value $x_t$ with positive density, $f(d_t \mid X_t=x_t) > 0$ over the range of doses considered. Finally, we assume that unobserved time-invariant heterogeneity $U$ affects treatment only through $X_t$, i.e., $D_t \indep U \mid X_t$. This assumption implies that, conditional on $X_t$, $U$ does not introduce additional confounding of the $Y_t$ and $D_t$ relationship. In our COVID-19 application, $U$ may represent underlying socioeconomic and occupational characteristics, such as the prevalence of essential workers with limited ability to work from home. These factors can contribute to higher case counts and may also drive stronger policy responses. Figure~\ref{dag} illustrates a general longitudinal causal structure consistent with these assumptions, allowing for treatment–confounder feedback and unobserved time-invariant heterogeneity and targeting a marginal estimand.

\begin{figure}
	\centering
	\begin{tikzpicture}[
		scale=0.9,
		every node/.style={
			circle,
			draw,
			minimum size=0.7cm,
			inner sep=1pt
		},
		>=stealth
		]

		\node (X) at (0,0) {$X_t$};
		\node (D) at (3,0) {$D_t$};
		\node (Y) at (1.5,2.5) {$Y_t$};
		\node (U) at (-0.5,2.5) {$U$};

		\draw[->] (X) -- (D);
		\draw[->] (X) -- (Y);
		\draw[->] (D) -- (Y);

		\draw[->, gray] (U) -- (X);
		\draw[->, gray] (U) -- (Y);
		
	\end{tikzpicture}
	\caption{Directed acyclic graph (DAG) for identification of the marginal dose--response function
		$\mu(d)=\mathbb{E}\{Y(d)\}$ under marginal ignorability. Observed covariates $X_t$ block all backdoor paths between exposure $D_t$ and outcome $Y_t$, while $D_t$ depends on unobserved time-invariant heterogeneity $U$ only through $X_t$.}
	\label{dag}
\end{figure}
Under these assumptions, the dose–response function  $\mu(d)$ is identified by
\begin{equation}
	\E[Y_{t}(d)] = \E_{X_t}\left[\E\left[Y_{t}(d)|X_t\right]\right] =\E_{X_t}\left[\E\left[Y_{t} \mid D_{t}=d,X_t,e_t(d, X_{t})\right]\right],
	\quad d \in \mathcal D,
	\label{eq:mu_identified}
\end{equation}
where $e_t(d, X_{t}) = f(D_t= d\mid X_t)$, following the balancing property of the GPS. Our target estimand is the marginal APO, summarized by the marginal structural model (MSM),
\[
g\left(\E\!\left\{ Y(d) \right\}\right) = m(d;\xi),
\]
where $g(\cdot)$ is a known link function and $\xi$ is defined through the marginal
dose–response curve.  Two causal estimation approaches can be considered. First, under GPS weighting, let $w_t=f(D_t = d)/{e_t(d,X_t)}$ denote the stabilized inverse GPS weight. For any fixed $d\in\mathcal D$,  the reweighted joint distribution satisfies
\[\frac{w_t\,f(D_t=d\mid X_t)\,f(X_t)}
{\int w_t\,f(D_t=d\mid x)\,f(x)\,dx}=
\frac{f(D_t=d)\,f(X_t)}{\int f(D_t=d)\,f(x)\,dx}=f(X_t),\]
that is, in the weighted pseudo-population, the distribution of covariates $X_t$ among $D_t=d$ equals the marginal distribution of $X_t$. Therefore, the weighted conditional expectation identifies the marginal APO
\begin{equation*}
	\E_w[Y_t\mid D_t=d]= \int \E[Y_t\mid D_t=d,X_t]f(X_t)\,dX_t = \mu(d).
	\label{eq:weighted_identification}
\end{equation*}
Let $\nu^{W}_{it}(\xi) = g^{-1}\{m(D_{it};\xi)\}$ and define $\nu^{W}_i(\xi) = (\nu^{W}_{i1}(\xi),\ldots,\nu^{W}_{iT_i}(\xi))^\top$. The MSM parameter $\xi$ is estimated by solving the GEE \citep{preisser2002performance}
\begin{equation}\label{gee}
	\sum_{i=1}^{n} \mathbf{U}_i^{W}(\xi) = 0,\;\;\; \mathbf{U}_i^{W}(\xi)
	=
	\left( \frac{\partial \nu^{W}_i(\xi)}{\partial \xi} \right)^\top
	\Sigma_i^{-1}
	w_i
	\left( Y_i - \nu^{W}_i(\xi) \right),
\end{equation}
where $w_{i}= diag(w_{i1},\ldots,w_{iT_i})$, $Y_i = (Y_{i1},\ldots,Y_{iT_i})^\top$, and $\Sigma_i$ is a $T_i \times T_i$ working covariance matrix. According to \cite{tchetgen2012specifying}, consistent estimation of the MSM parameter $\xi$ is guaranteed under an independence working correlation. Further technical details are provided in Appendix~\ref{app:canonical}.

Second, under GPS regression adjustment \citep{hirano2004propensity}, we can specify the working conditional outcome model as
\begin{equation*}
	g\left\{\E[Y_t \mid D_t, X_t, \hat e_t(X_t); \tau]\right\}= s_1(X_t,\beta)+h(D_t,\hat e_t(X_t),\theta),
\end{equation*}
where $s_1(\cdot)$ and $h(\cdot)$ are specified functions, $\tau = (\beta^\top,\theta^\top)^\top$, and $\hat e_t(X_{it})$ is the estimated GPS. In this working model, the parameter $\tau$ is estimated by solving the GEE,
\begin{equation}\label{gee1}
	\sum_{i=1}^{n} \mathbf{U}_i^{R}(\tau) = 0, \;\;\; \mathbf{U}_i^{R}(\tau)
	=
	\left( \frac{\partial \nu^{R}_i(\tau)}{\partial \tau} \right)^\top
	\Sigma_i^{-1}
	\left( Y_i - \nu^{R}_i(\tau) \right),
\end{equation}
with $\nu^{R}_i(\tau) = \big(\nu^{R}_{i1}(\tau), \ldots, \nu^{R}_{iT_i}(\tau)\big)^\top$ and $\nu^{R}_{it}(\tau)=g^{-1}\!\left\{s_1(X_{it};\beta) + h(D_{it},\hat e_t(X_{it});\theta)\right\}$. Under this working model, the marginal APO is identified via regression standardization, 
\begin{equation*}
	\mu_t(d)=
	\E_{X_t}[
	g^{-1}(
	s_1(X_t,\beta)+h(d,e_t(X_t),\theta))
	].
	\label{eq:mu_standardized}
\end{equation*}
In practice, after fitting the working conditional model, we estimate $\mu_t(d)$ by
\begin{equation*}
	\hat\mu_t(d)=\frac{1}{n}\sum_{i=1}^ng^{-1}(s_1(X_{it},\hat \beta)+h(d,\hat e_t(X_{it}),\hat \theta)).
	\label{eq:mu_hat_standardized}
\end{equation*}
In this unweighted setting, the GEE parameter is consistent given a correctly specified conditional mean model regardless of whether the working correlation structure is correct  \citep{tchetgen2012specifying}. This GEE framework offers flexibility in targeting the marginal treatment effect for repeated measurements. However, its reliance on estimating equations, rather than full likelihoods, poses challenges for conventional Bayesian analysis, as there is no explicit distributional form for the marginal mean model. This motivates the development of a Bayesian approach, which we outline in the next section.

\subsection{Bayesian non-parametric method for GEE-based causal estimands}
\label{Sec:bb}
The Bayesian framework offers natural uncertainty quantification and prediction, which can provide a more thorough understanding of the causal effect from a practical perspective. There is growing interest in the application of Bayesian methodology to causal problems, but most proposed methods appeal to the principles of propensity score-adjustment or flexible outcome modeling \citep{stephens2021bayesian, li2023bayesian}. An advantage of the Bayesian approach for estimation of these models is that estimated causal effects are in the form of distributions, rather than point estimates, allowing us to make probability statements about the causal quantities of interest (i.e., APOs). In fully parametric Bayesian analyses, this is typically achieved by specifying likelihood models for both the outcome and treatment assignment mechanisms.  However, such an approach is not directly applicable in the present setting, as the GEE in \eqref{gee} or \eqref{gee1} specifies only the marginal or conditional mean structure of the outcome and does not correspond to a fully specified likelihood. Consequently, Bayesian inference in this context must proceed under an approximate (or partially specified) model, which may be misspecified relative to the true data-generating mechanism. Rather than modeling the full joint distribution of the data, inference targets the population root of the unbiased estimating equation used to identify a causal estimand. This perspective aligns with the generalized Bayesian framework and recent work on Bayesian inference for loss functions \citep{luo2023assessing}. 

Although the generalized Bayesian posterior in this section is written using expectations under observational setting, our inferential target is causal and is defined under a hypothetical intervention on the dose. Following the Bayesian causal framework in \citet{liu2020estimation}, let $\mathcal{O}$ denote the observational law of $Z=(Y,X,D)$ and note that $\mathcal{O}\equiv F_0$, where $F_0$ denotes the true, unknown data-generating distribution. For each dose level $d\in\mathcal D$, let $\mathcal{E}_d$ denote the law of $Z$ under the hypothetical intervention that sets the dose to the constant level $d$ at each time point (equivalently, the law governing $Y(d)$ under SUTVA and consistency as defined in Section~\ref{sec:apo}). Our causal estimand is $\mu(d)=\mathbb{E}_{\mathcal{E}_d}(Y)$ (or $\mu_t(d)=\mathbb{E}_{\mathcal{E}_d}(Y_t)$), and the causal MSM parameter $\xi^\star$ is defined by
\[
g\{\mu(d)\}=m(d;\xi^\star),\qquad d\in\mathcal D.
\]
Under the identification assumptions in Section~\ref{sec:apo}, $\mu(d)$ (and hence $\xi^\star$) is identified as a functional of the observational law $\mathcal{O}$. Therefore, the estimating equation $\mathbf U^{W}(\xi)$ below is constructed from the identified observed-data representation of $\mu(d)$ so that its population root coincides with the causal MSM parameter $\xi^\star$.
Specifically, we define a generic GEE $\mathbf{U}(\psi)$, which may represent either the GPS-weighted GEE or the regression-adjusted GEE, depending on the parameter of interest:
\[
\mathbf{U}(\psi) =
\begin{cases}
	\mathbf{U}^{W}(\xi), & \text{if $\psi = \xi$ (MSM parameter)}, \\
	\mathbf{U}^{R}(\tau), & \text{if $\psi = \tau$ (working regression parameter)}.
\end{cases}
\]
$\mathbf U^{W}(\xi)$ targets the causal MSM parameter $\xi^\star$ (hence $\mu(d)$), whereas $\mathbf U^{R}(\tau)$ targets a working regression parameter used for regression standardization; the causal estimand remains $\mu(d)$ rather than $\tau$ itself (see Appendix~\ref{app:iden} for additional discussion).

At the population level, the parameter $\psi_0$ is defined as the solution to the population moment condition
\begin{equation}
	\label{eq:population_moment}
	\mathbb{E}_{F_0}\big[\mathbf{U}(\psi_0)\big] = 0.
\end{equation}
Equivalently, $\psi_0$ can be characterized as the minimizer of an expected loss function
\begin{equation}
	\label{eq:population_loss}
	\psi_0 = \arg\min_{\psi \in \Psi} \; \mathbb{E}_{F_0}\big\{ \ell(Z;\psi) \big\},
\end{equation}
where $\ell(Z;\psi)$ is a loss function corresponding to $\mathbf{U}(\psi)$, i.e., $\mathbf{U}(\psi) = \partial \ell(Z;\psi)/\partial \psi$. This loss-based definition is consistent with the generalized Bayesian framework of \cite{luo2023assessing} and defines the population root of the moment condition used to identify the target causal estimand without requiring a likelihood. In the Bayesian non-parametric framework, uncertainty about $F_0$ is represented through a posterior distribution over probability measures. If we assume that the data points are realizations from a multinomial model on the finite set of the observed data $\left\{z_1,\ldots,z_n\right\}$ with unknown probability $\mathbf{p}=\left(p_1,\ldots,p_n\right)$, and assuming a priori that $\mathbf{p}\sim \text{Dirichlet}\left(\alpha,\ldots,\alpha\right)$, then, a posteriori $\mathbf{p}\sim \text{Dirichlet}\left(\alpha+1,\ldots,\alpha+1\right)$. The standard Bayesian bootstrap is obtained under the improper specification $\alpha=0$ \citep{rubin1981bayesian}.  Therefore, if we repeatedly sample $\mathbf{p} \sim  \text{Dirichlet}\left(1,\ldots,1\right)$, then 
\[
F^{(b)} = \sum_{i=1}^n p_i^{(b)} \delta_{Z_i}.
\]
For each posterior draw $F^{(b)}$, the corresponding parameter draw $\psi^{(b)}$ is defined as the solution to the posterior predictive estimating equation
\begin{equation}
	\label{eq:bb_estimating_equation}
	\int \mathbf{U}(\psi^{(b)}) \, dF^{(b)}(z)
	=
	\sum_{i=1}^n p_i^{(b)} \mathbf{U}_i(\psi^{(b)}) = 0.
\end{equation}
This construction induces a posterior distribution over $\psi$ through the posterior over $F$, yielding valid Bayesian uncertainty quantification for GEE-based causal estimands \citep{newton1994approximate,chamberlain2003nonparametric}. Importantly, the definition of $\psi_0$ in \eqref{eq:population_moment} and the Bayesian bootstrap implementation in \eqref{eq:bb_estimating_equation} represent equivalent formulations of the same inferential target. The former defines the population root of the GEE to identify the causal estimand, while the latter provides a practical posterior sampling scheme. This approach yields a coherent Bayesian non-parametric framework for inference on GEE-based parameters, accommodates model misspecification, and avoids the need to specify a full likelihood while retaining the interpretability of marginal causal estimands.

\subsection{The generalized Bayesian bootstrap}
The previous section describes a standard Bayesian bootstrap approach for obtaining a posterior distribution for parameters defined through GEEs. In our setting, the model is specified through a two–stage system of GEEs, consisting of a GPS model in the first stage and an outcome model in the second stage. Specifically, let $\mathbf{U}_1(\gamma)$, parameterized by $\gamma$, denote the GEE associated with the GPS model and $\mathbf{U}_2(\psi)$ with parameter $\psi$, denote the GEE which may correspond to either the GPS-weighted estimating equation in \eqref{gee} or the regression-adjusted estimating equation in \eqref{gee1}. Under the Bayesian bootstrap, we repeatedly sample weights $\mathbf{p} = (p_1,\ldots,p_n) \sim \mathrm{Dirichlet}(1,\ldots,1)$, and for each draw of $\mathbf{p}$, solve the weighted estimating equations
\begin{align}
	\sum_{j=1}^{n} p_j \, \mathbf{U}_1(x_j, d_j, \gamma) &= \mathbf{0}, \label{b1} \\
	\sum_{j=1}^{n} p_j \, \mathbf{U}_2\!\left(z_j, e(x_j;\hat\gamma),\psi\right) &= \mathbf{0}, \nonumber
\end{align}
where $e(x_j;\gamma)$ denotes the GPS model, and $\hat\gamma$ is the solution to \eqref{b1}. The resulting solution $\hat\psi(\mathbf{p})$ defines a draw from the posterior distribution of $\psi$ induced by the Bayesian bootstrap. Repeating this procedure yields an empirical approximation to the posterior distribution of $\psi$. While this Bayesian bootstrap specification provides a convenient and likelihood-free way to propagate uncertainty through the two-stage GEE system, it also has an important limitation. The Bayesian bootstrap places all its prior mass on the empirical support of the observed data, implicitly assuming that the unknown data-generating distribution $F$ is discrete with atoms fixed only at the observed points. As a result, posterior inference for $\mu(d)$ is entirely driven by reweighting the observed sample and cannot extrapolate beyond the empirical support of the observed data. It often tends to underestimate uncertainty in regions where data are sparse, particularly in the tails of the distribution.

As proposed by \cite{luo2023assessing}, adopting a DP prior with base measure $G_0$ and concentration parameter $\alpha > 0$ allows posterior draws of the data-generating distribution to place mass both on the observed data and on new atoms drawn from $G_0$. Consequently, posterior predictive distributions for the marginal dose-response function $\mu(d)$ are no longer confined to the empirical support of the observed doses. This enables uncertainty propagation in regions of the dose space that are sparsely observed or lie near the extremes of the exposure distribution. Moreover, the base measure $G_0$ can carry weak structural information and can be interpreted as prior knowledge about the underlying data-generating process. In addition, the parameter $\alpha$ and $G_0$ can serve as a regularization mechanism, stabilizing posterior inference in regions with sparse data by shrinking toward the base measure while still allowing the data to dominate where information is abundant.

Therefore, we extend the Bayesian bootstrap to a more general DP specification. If we specify a priori $P \sim DP\left(\alpha,G_0\right)$ where $\alpha>0$ is the concentration parameter and $G_0$ is the base measure. Given data $\left(z_1,\ldots,z_n\right)$, the resulting posterior distribution of $P$ is $DP\left(\alpha_n,G_n\right)$, where $\alpha_n=\alpha+n$ and $G_n(\ldotp)= \alpha G_0(\ldotp) \big /{(\alpha+n)} + {\sum_{j=1}^{n}\delta_{z_j}\left(\cdot\right)}\big/{(\alpha+n)}$.  Specifically, we can generate the collection of data from the DP via $\left\{z_j\right\}_{j=1}^{\infty} \sim G_n$ and $\left\{p_k\right\}_{j=1}^{\infty} \sim StickBreaking(\alpha_n)$; the standard stick-breaking algorithm \citep{sethuraman1994constructive} generates the weights by a transformation of the collection $\{V_k\}_{k=1}^\infty$ where $V_j \sim Beta(1,\alpha)$ are independent, with $p_1 = V_1$ and for $j=2,3,\ldots$, $p_j = V_j \prod_{k=1}^{j-1} (1-V_k)$. Therefore, to obtain a posterior variate, we can sample $\left\{z_j^{s}\right\}_{j=1}^{\infty} \sim G_n$ and $\left\{p_j\right\}_{j=1}^{\infty} \sim StickBreaking(\alpha_n)$, then solve
\begin{equation}
	\label{a1}
	\sum_{j=1}^{\infty} p_j\mathbf{U}_1\left(x^{s}_j,d^{s}_j,\gamma\right)= \mathbf{0} ,\sum_{j=1}^{\infty} p_j\mathbf{U}_2\left(z^{s}_j,e(x_j, \hat \gamma),\psi\right)= \mathbf{0}.
\end{equation}
Although this is an infinite sum, the $p_j$ decreases in expectation as $j$ increases and eventually becomes numerically negligible. Because of this decay, the infinite sum can be truncated at some finite 
$J$, such that $p_J <\epsilon$ for a small tolerance $\epsilon$.  The uncertainty about $\psi$ is properly addressed in this `forward' calculation. We detail this DP specification for the longitudinal analysis in Algorithm \ref{A0}. In this formulation, we do not directly specify a prior for $\psi$ but instead it is a functional of $P$, where we carry out a Bayesian inference for $P$. This approach is computationally more efficient because it relies purely on optimization instead of MCMC, which provides a ‘shortcut’ to Bayesian causal inference. The validity and theoretical justification of this approach have been thoroughly examined in cross-sectional settings in \cite{luo2023assessing}.

\begin{algorithm}[ht]
	\SetAlgoLined
	\KwData{$z_{1:n}=(z_1,\ldots,z_n)$}
	Given $\alpha$ and $G_0$,
	\For{$s = 1$ \textbf{to} $S$}{
		\begin{itemize}
			\item 
			Sample a new collection of data $\{x^{s}_j,d^{s}_j\}_{j=1}^{J}$ using simple random sampling with replacement from the original dataset. For each sampled unit, generate $y^{s}_j$ from the existing observations $\{y_i\}_{i=1}^{n}$ with probability $1/(\alpha+n)$ or from $G_0$ with probability $\alpha/(\alpha+n)$;
			\item 	Sample $\{p_j^{s}\}_{j=1}^{J}$ from a stick-breaking process with $\alpha_n=\alpha+n$; 
			
			\item     Estimate the GPS based the GEE related to $\mathbf{U}_1$ in \eqref{a1} with the data  $\{x^{s}_j,d^{s}_j\}_{j=1}^{J}$ and the weight $\{p_j^{s}\}_{j=1}^{J}$; 
			
			\item 	Solve the GEE related to $\mathbf{U}_2$ in \eqref{a1} and obtain $\psi^{s}$ with the data $z^{s}= \{y^{s}_j,x^{s}_j,d^{s}_j\}_{j=1}^{J}$, the weight $\{p_j^{s}\}_{j=1}^{J}$ and the estimated GPS;
			
			\item    Compute the APO, $\mu^s(d)$, at a given dose level $d$ based on the sample $\psi^{s}$.
		\end{itemize}
	}
	\Return $(\mu^1(d),\ldots,\mu^S(d))$. \\[6pt]
	\caption{\label{A0}Algorithm for generating posterior samples of the APO at a given dose level $d$ based on the generalized Bayesian bootstrap (DP specification).}
\end{algorithm}

In this setting, once a new dataset is generated, we first estimate the parameters based on the GEE related to $\mathbf{U}_1$, and then plug the estimated GPS into the GEE associated with $\mathbf{U}_2$. This procedure induces a random empirical measure that weights both the GPS and outcome estimating equations (with the same resampling weights) so that uncertainty is propagated coherently through the two-stage system. Both the GPS model and the outcome model are specified non-parametrically and weighted with the same stick-breaking weight. This distinguishes our approach from the previous work in \cite{saarela2015bayesian,liu2020estimation}, where the weight is only placed in the outcome model.  This is known as a two-step procedure \citep{kaplan2012two} and is commonly used in Bayesian causal inference to avoid model feedback, as GPSs should be estimated without incorporating outcome information in order to preserve their balancing properties \citep{rubin2007design,mccandless2010cutting,saarela2015bayesian}. This procedure has been shown to yield preferable estimation performance in practice \citep{stephens2021bayesian}. \cite{stephens2021bayesian} further showed that weighting both models with the same Bayesian bootstrap weights, analogous to our approach of using a common stick-breaking weight, yields valid coverage. In addition, our Bayesian implementation in the longitudinal setting performs resampling at the subject level rather than at the observation level. Specifically, we assign a single stick-breaking weight to each study unit, which is held fixed across time. When a study unit is resampled, their entire longitudinal trajectory $\{D_{it}, X_{it}, Y_{it}\}_{t=1}^T$ is included or generated from the base measure in the resampled dataset. This subject-level resampling strategy preserves the within-subject dependence structure across repeated measurements and aligns with the marginal modeling perspective. By contrast, resampling study unit observations would implicitly impose independence across time within subjects and would be incompatible with the longitudinal dependence encoded in the GEE. 

In terms of GEE-based identification under this framework, \citet{luo2023assessing} show that, under suitable regularity conditions, the generalized Bayesian  bootstrap procedure built upon GEE or more general loss functions yield posterior distributions that concentrate asymptotically around the corresponding frequentist $M$-estimators. In particular, when the treatment model is correctly specified, the estimated GPS provides valid adjustment for treatment assignment the GEE, allowing the weighted GEE to yield unbiased estimating equations for the target marginal parameters. Consequently, the resulting posterior for the APO is supported on the identified parameter space characterized by \eqref{eq:mu_identified} and concentrates around the corresponding frequentist estimand, which coincides with the true marginal dose–response function under identification assumptions stated in Section \ref{sec:apo}.

\section{Simulations}
\label{sec:sim}
In this section, we examine the performance of the methods described in Section \ref{sec:meth} in different aspects.  The code for the simulation is publicly available on our GitHub page at   \url{https://github.com/yumcgill/Bayes_DoseResponse}. We consider the following models for our simulation studies.
\begin{itemize}
	\item \textbf{WOR-BB:} Marginal APO estimated using the GPS weighting. The posterior distribution of the associated parameters is obtained via \eqref{b1} with $\mathbf{U}_2 = \mathbf{U}^{W}(\xi)$. The marginal treatment density $f(d)$ is estimated nonparametrically using a kernel density estimator.
	\item \textbf{COV-BB:} Marginal APO estimated using GPS regression adjustment. The posterior distribution of the associated parameters is obtained via \eqref{b1} with $\mathbf{U}_2 = \mathbf{U}^{R}(\tau)$.
	\item \textbf{WOR-DP:} Marginal APO estimated using GPS weighting. The posterior distribution of the associated parameters is obtained via \eqref{a1} with concentration parameter $\alpha=5$ (Algorithm \ref{A0}), using $\mathbf{U}_2 = \mathbf{U}^{W}(\xi)$. The marginal treatment density $f(d)$ is estimated nonparametrically using a kernel density estimator.
	\item \textbf{COV-DP:} Marginal APO estimated using GPS regression adjustment. The posterior distribution of the associated parameters is obtained via \eqref{a1} with concentration parameter $\alpha=5$ (Algorithm \ref{A0}), using $\mathbf{U}_2 = \mathbf{U}^{R}(\tau)$.
	\item \textbf{Bayes-MSM}: A Bayesian MSM in which the causal dose–response function is modeled directly as a parametric function of the dose. Stabilized inverse GPS weights, $w_t = f(d)/\hat e_t(d, X_t)$, are incorporated through a weighted likelihood to remove time-dependent confounding. The GPS and outcome models  are both modeled parametrically but completely separately. The outcome model is specified as a cubic spline regression of the outcome on dose, with non-informative priors placed on all regression coefficients and variance parameters. Posterior inference for both GPS and outcome models is obtained via MCMC using \texttt{Stan} \citep{stanref}. Note this formulation is not a Bayesian analysis based on a fully specified joint likelihood factorization for all observed processes with independent priors on treatment and outcome model parameters \citep[see discussion in][]{saarela2015bayesian}; rather, it follows a Bayesian working-model approach in which weighting is used to target the causal estimand directly.
\end{itemize}

Specifically, we outline the procedure for generating posterior samples of the APO at a given dose level $d$ using the proposed DP–based method:
\begin{enumerate}
	\item \textbf{Resampling under the DP prior.}
	For each Monte Carlo iteration $s = 1,\ldots,S$, we first resample with replacement from the original dataset to obtain a new dataset with the number of study units $N=100$. This step can be viewed as simple random sampling with replacement. For each resampled study unit, a random draw determines whether the study unit’s outcome, $y$, is taken directly from the base distribution, with probabilities $\alpha/(\alpha+n)$. For study units drawn from the base distribution, new outcome values are simulated from a normal distribution with mean given by the model-based predicted value and a standard deviation of 0.5. For each study unit, the predicted mean is derived from the outcome model specified in the corresponding simulation example.
	\item \textbf{Weight generation via stick-breaking.}
	Given the resampled data, a corresponding set of weights $\{p_j^{(s)}\}_{j=1}^{100}$ is drawn from a stick-breaking process with concentration parameter $\alpha_n = \alpha + n$. These weights are assigned to the resampled study units and carried forward into the analysis. 
	\item \textbf{GPS estimation (first stage).}
	Using the resampled data from Step 1  and weights from Step 2, the GPS is estimated by solving GEE associated with $\mathbf{U}_1$. This stage is performed independently of the outcome model in order to avoid feedback from the outcome into the GPS estimation.
	\item \textbf{Outcome model estimation (second stage).}
	Conditional on the estimated GPS, the GEE associated with $\mathbf{U}_2$  is solved using the same resampled data and weights, yielding a posterior draw $\psi^{(s)}$ of the working parameter.
	\item \textbf{Causal estimand computation.}
	Based on the parameter draw $\psi^{(s)}$ and the estimated GPS, the posterior draw of the marginal APO at a specific dose $d$ is 
	\[
	\mu^{(s)}(d) =
	\begin{cases}
		g^{-1}\left[m(d;\xi^{(s)})\right], & \text{WOR},\\
		\E_{X}\big[ g^{-1}\{ s_1(X;\beta^{(s)}) + h(d,\hat e(X);\theta^{(s)}) \} \big], & \text{COV}.
	\end{cases}
	\]
	Repeating this procedure for $s = 1,\ldots,S$ produces posterior samples
	\[
	\{\mu^{(1)}(d), \ldots, \mu^{(S)}(d)\},
	\]
	which can be summarized to quantify uncertainty in the marginal APO via posterior means, credible intervals, and coverage rates.
\end{enumerate}

\subsection{Example 1}
In this example, we conduct a simulation study on a sample size of 100 study units, each with 10 time points and a total of 1000 observations. We generate the data via following  relationships:
\begin{equation*}
	\begin{aligned}
		U_{i}  \sim\mathcal{N} \left(0.2,0.1\right), X_{1it}  &\sim\mathcal{N} \left(0.2,0.1\right) + 0.25U_i , X_{2it}  \sim\mathcal{N} \left(1,0.6\right)\\
		D_{it} | X_{1it}, X_{2i}&\sim \mathcal{N} (5+4 X_{1it} +2 X_{2i} , 1) \\
		Y_{it}|D_{it}, X_{1it}, X_{2i},U_{i} &\sim \mathcal{N} \left(20\exp\left[D_{it}+X_{1it} - 0.25 X_{2it}+0.5U_i\right],1\right) \\
	\end{aligned}
\end{equation*}
$D$ is a continuous treatment variable, and $X_{1}$, $X_2$ are time-varying confounders and $U$ is a time-invariant heterogeneity with a nonlinear relationship in $Y$. The goal of this causal setting is to estimate the APO, i.e., the outcome at a specific dose level. For the GEE associated with the first stage of the GPS estimation, $\mathbf{U}_1$, we include only the covariates $X_{1it}$ and $X_{2it}$. For the GEE associated with the outcome model, $\mathbf{U}_2$, we use the outcome on log scale, $\log Y_i$.  Specifically, in the covariate-adjusted (COV) method, we include $D_{it}$ and a cubic spline of the estimated GPS, allowing for a flexible, nonlinear relationship between the GPS and the outcome.  In the weighted outcome regression (WOR) method, we include a cubic spline of $D_{it}$ only. For the DP-based method, the predicted mean is obtained from a GEE outcome model that includes $D_{it}$ and a cubic spline of the estimated GPS.

\begin{figure}[ht]
	\centering
	\includegraphics[width=1\linewidth]{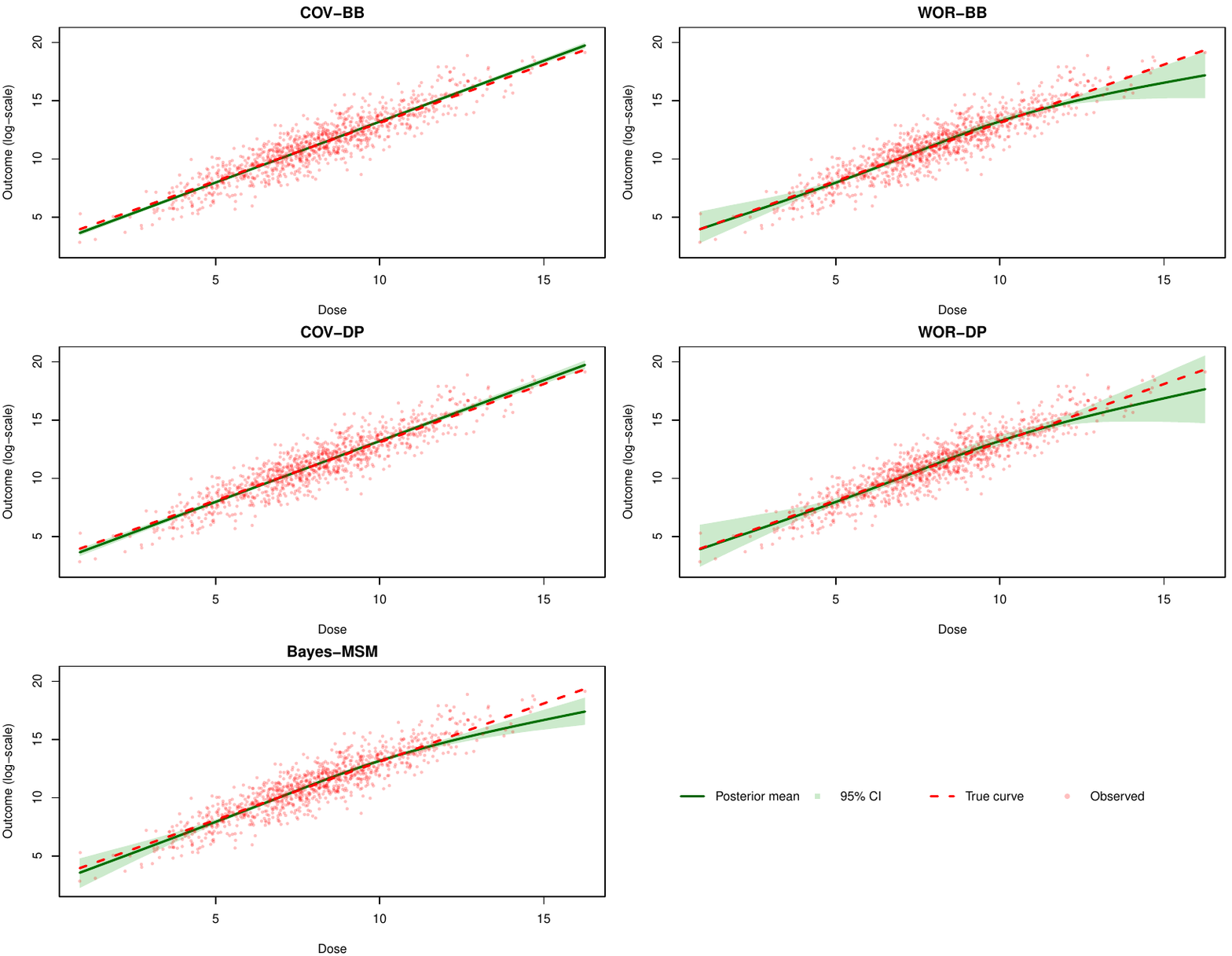}
	\caption{Example 1: Posterior predictive dose–response curves for the Gaussian outcome estimated using the Bayesian bootstrap, Dirichlet process, and Bayesian parametric MSM approaches.}
	\label{ex1}
\end{figure}

We first simulate one dataset and generate 1000 posterior samples. Figure \ref{ex1} shows the 1000 posterior dose-response curves with observed data points. All methods recover the general monotonic increasing relationship between dose and outcome, but there are differences in tail behavior and uncertainty quantification. The WOR-based approaches exhibit some deviation from the true dose-response curve, particularly in the high dose region. However, these regions are characterized by larger variances, leading to wider credible intervals that maintain coverage of the true curve. The DP-WOR method yields slightly wider intervals than BB-WOR, reflecting the additional uncertainty introduced by the more flexible prior on the data-generating distribution. In contrast, the COV-based approaches produce smoother dose-response estimates with relatively narrow credible intervals and remain close to the true curve even in regions with limited data support. This suggests that the COV-based formulation provides a more stable and reliable estimation framework, particularly in settings with small sample sizes. Finally, the Bayes-MSM approach also yields smooth estimates with comparatively tight credible intervals, but shows noticeable departures from the true curve in both tails, highlighting its sensitivity to parametric assumptions when extrapolating beyond well-supported regions of the data.

Subsequently, we repeat this simulation on 1000 datasets, and evaluate the performance on three given doses. The results are shown in Table \ref{long}, with averages of the posterior means (Est), standard deviations (SD) and coverage rates based on the 2.5\% and 97.5\% posterior sample quantiles. The true APO at dose $d$ was estimated by Monte Carlo integration with 100,000 observations, corresponding to 10,000 study units with 10 measurements each. The APO estimates are calculated for doses 6, 8 and 10. The APO estimates in all models are represented via a logarithmic transformation, and the results are also presented in the logarithmic scale.  All methods yield the average of the posterior means close to the true values. The COV-based approaches are  more efficient than the WOR-based ones, with smaller standard deviations. The DP-based methods attain near-nominal coverage across all doses considered, whereas COV-BB achieves nominal coverage only at doses with adequate data support and exhibits undercoverage at sparsely observed doses. The WOR-BB method shows consistently lower coverage across doses, while the Bayes-MSM displays undercoverage at doses with limited data support. Overall, this example indicates that the proposed DP-based methods are well suited for longitudinal data with repeated measures, providing valid uncertainty quantification even in low-data regions.

\begin{table}[ht]
	\caption{\label{long} Simulation results for the dose–response function at three dose levels, reporting the average posterior mean (Est) with associated standard deviation (SD), and empirical coverage rate (Coverage, \%). Results are summarized over 1,000 simulation replicates.}
	\centering
	\scalebox{0.95}{\begin{tabular*}{30pc}{@{\hskip5pt}@{\extracolsep{\fill}}c@{}c@{}c@{}c@{}c@{\hskip5pt}}
			\multicolumn{5}{l}{Example 1 (Results are presented in the logarithmic scale)}\\
			\hline
			Method	&  & dose = 6 &  dose = 8& dose = 10	\\
			& & (9.092) & (11.092) & (13.092)\\
			\hline
			COV-BB &Est& 9.028&  11.096 & 13.163\\
			& SD& 0.049&  0.040 &  0.050\\
			& Coverage & 76.2 &  95.2 & 68.8\\ 
			\hline
			WOR-BB &Est&9.071 & 11.094 &  13.118 \\
			& SD & 0.213 &  0.160 & 0.222\\
			& Coverage & 86.4 & 83.6 & 81.7\\ 
			\hline
			COV-DP &Est&9.030 &  11.097 &  13.165 \\
			& SD & 0.092 &  0.061& 0.093\\
			& Coverage & 98.2 & 99.0 & 96.5 \\ 
			\hline
			WOR-DP & Est&9.065 &  11.094&  13.123 \\
			& SD & 0.170&  0.131&  0.181\\
			& Coverage & 97.4 & 98.3 &  97.7 \\ 
			\hline
			Bayes-MSM & Est&9.064 & 11.097& 13.129 \\
			& SD &0.101& 0.070 & 0.106 \\
			& Coverage & 88.1 & 93.3 & 85.5\\ 
			\hline
			\multicolumn{5}{l}{Example 2}\\
			\hline
			Method	&  & dose = 6 &  dose = 8& dose = 10		\\
			& & (9.207) &  (13.735)  & (20.490)\\
			\hline
			COV-BB & Est&9.180&  13.744 & 20.578\\
			& SD & 0.132 &  0.132 &  0.191\\
			& Coverage & 93.4 & 93.4 & 93.0\\ 
			\hline
			WOR-BB & Est&9.189 & 13.735 &  20.520 \\
			& SD &0.571 &  0.532 &  0.931\\
			& Coverage & 85.2 & 84.9 & 82.7\\ 
			\hline
			COV-DP & Est& 8.852 &  13.219 &  19.752 \\
			& SD & 0.213&  0.194 & 0.385\\
			& Coverage & 97.2 & 91.9 & 97.1\\ 
			\hline
			WOR-DP & Est&8.850 &  13.213 & 19.692\\
			& SD & 0.452& 0.429 & 0.752\\
			& Coverage &  95.8 &  94.5 &  94.2 \\ 
			
			\hline
			Bayes-MSM & Est&9.152 &  13.687& 20.450 \\
			& SD &0.287& 0.264& 0.432 \\
			& Coverage & 85.6 & 90.1 & 91.6\\ 
				\hline
		\end{tabular*}
	}
\end{table}

\subsection{Example 2}
In this example, we demonstrate inference for simulated outcome as the count data, similar to our real data. All the variables simulated from the same distribution as in Example 1, except the outcome which is generated from a Poisson distribution:  
\[
Y_{it}|D_{it}, X_{1it}, X_{2i},U_{i} \sim \text{Poisson} \left(\exp\left[1+0.2 D_{it}+0.005 \frac{X_{1it}}{100} - 0.002 \frac{X_{2it}}{100}+0.1U_i\right]\right). 
\]

Similar to the first example, {for the GEE associated with the first stage of GPS estimation, we include only the covariates $X_{1it}$ and $X_{2it}$. For the outcome model, we fit a Poisson GEE for $Y_i$. 
	In the COV method, the model includes $D_{1it}$ and a cubic spline of the estimated GPS. In the WOR method, the outcome mean is modeled using a cubic spline of $D_{1it}$, with the estimated GPS incorporated as a weight. For the DP-based method, the predicted mean is obtained from a GEE outcome model including $D_{it}$ and a cubic spline of the estimated GPS.}

\begin{figure}[ht]
	\centering
	\includegraphics[width=1\linewidth]{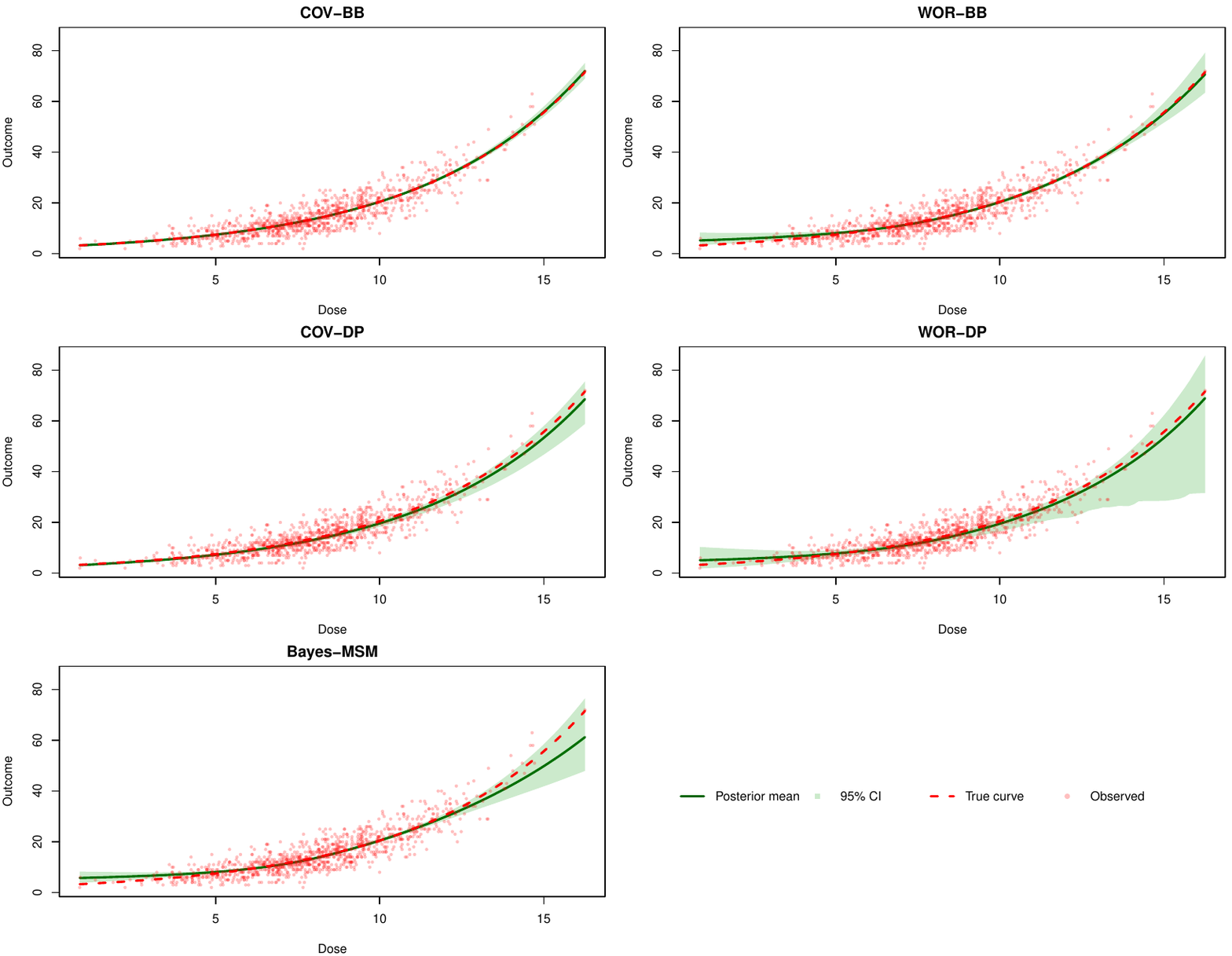}
	\caption{Example 2: Posterior predictive dose–response curves for the Poisson outcome estimated using the Bayesian bootstrap, Dirichlet process, and Bayesian parametric MSM approaches.}
	\label{ex2}
\end{figure}

Figure \ref{ex2} demonstrates the 1000 posterior dose-response curves with observed data points. Both BB-based and DP-based approaches closely align with the true dose-response curve, with only small deviations at the high-dose tail. The WOR-based formulations exhibit wider 95\% credible intervals, reflecting increased uncertainty in regions with limited data support. Similar with the previous example, the DP-based methods produce slightly wider intervals at extreme doses, capturing additional uncertainty induced by the more flexible prior. The Bayes-MSM method shows the most deviation of the posterior mean from the true curve at high doses, suggesting that the parametric-based MSM can lead to certain bias in regions where data are sparse. 

Table \ref{long} summarizes the results of 1,000 simulation runs at three specified dose levels. All methods give roughly unbiased estimates for all dose levels. WOR-based methods again exhibit larger standard deviations, with WOR-DP achieving near-nominal coverage, while WOR-BB shows systematic undercoverage across all doses. The COV-based methods maintain good coverage at all dose levels. Similar to the previous example, Bayes-MSM shows undercoverage at the dose extremes, highlighting its sensitivity in regions with limited data support as we observed in Figure \ref{ex2}.

\section{Application: Quantifying the causal impact of mass transit on COVID-19 transmission} 
\label{sec:real}
As discussed in the introduction, empirical studies quantifying the causal impact of mass public transport on COVID-19 transmission are scarce. For estimation, we have monthly (longitudinal) city level data available which measure COVID-19 cases, mass transit ridership, mobility containment measures, and a set of mobility characteristics for the period between 1 March 2020 and 23 January 2022. Longitudinal data provide the temporal resolution needed to understand how public transit ridership and COVID-19 incidence evolve and intersect.  Most previous studies use aggregate measures of mobility from mobile phone data sources such as those that were made publicly available by Google Mobility and Apple Maps during the pandemic, and issues of causality have received little attention. In our review, we have identified just one empirical study by \cite{Garcia2023} which uses causal analysis methods to assess the impact of travel specifically on metro networks (referred to as subways in the paper) on the transmission of COVID-19 in Valencia, Spain. In the following sections, we provide a summary of empirical studies undertaken on the impact of mobility on COVID-19 transmission, beginning with an overview of literature using non-causal analysis methods, followed by a summary of the limited literature adopting causal methods.

\subsection{Non-causal empirical studies}
There is a vast number of empirical studies that use non-causal statistical analysis methods to study the association of mobility with COVID-19 transmission as measured by either the number of COVID-19 cases or the virus reproduction number. Generally, most studies conclude that an increase in mobility is positively associated with an increase in COVID-19 transmission, and restrictions on mobility are associated with reductions in COVID-19 transmission.  Studies using statistical correlation methods are prevalent in the literature. For example, \cite{Jia2020} reported a positive linear correlation between mobility and case incidence in China at the early stages of the pandemic. \cite{Badr2020} analyzed mobility patterns in the USA and also reported positive linear correlations between mobility and COVID-19 case rates. \cite{Kissler2020} used Pearson's linear correlation coefficient and reported that their estimates of COVID-19 prevalence were negatively associated with reductions in mobility in their analysis of New York City. 

Studies adopting different forms of multivariate regression methods are also prevalent, and here we present a selection of examples. \cite{Li2021lancet} used multivariate linear regression and reported that the reproduction number of the virus increased with higher levels of mobility at retail and recreation places, workplaces, and transit stations in the UK. \cite{Kraemer2020} applied a generalized linear regression model framework and found that mobility was positively correlated with case incidence in China prior to travel restrictions, but the relationship was no longer statistically significant after the implementation of restrictions. \cite{McGrail2020} undertook a cross-country analysis of 134 countries and 50 US states. The authors used a generalized linear mixed effects model and reported that mobility and infection rates reduced after the imposition of social distancing restrictions. \cite{Islam2020} also performed a cross-country analysis of 149 countries using a two-stage linear regression analysis. It was reported that case incidence reduced with the imposition of public transport closures, school closures, work closures, restrictions on mass gatherings, and blanket lockdowns. \cite{Xiong2020} reported a positive relationship between mobility and COVID-19 cases in the US using structural equation modeling with dynamic panel and time-varying effects at the level of US counties. \cite{Manzira2022} used a multivariate time series regression model with autoregressive lag effects and reported a positive association between mobility and the incidence of COVID-19 in the city center areas of Dublin, Ireland. 

\subsection{Causal empirical studies}

There are a limited number of studies that adopt causal analysis methods to quantify the impact of mobility on COVID-19 case transmission. \cite{Garcia2023} undertook the only known analysis of the impact of metro travel on COVID-19 case rates. \cite{Garcia2023} performed a small-scale analysis of the impact of the locations of subway stations and hospitals on COVID-19 cases in Valencia, Spain. They first partitioned the region using Voronoi diagrams using subway locations and hospitals as segmentation variables, and then applied the Granger causality test. A positive association was reported between interchange subway stations and COVID-19 cases, while transmission from hospital locations generated more mixed results.

\cite{Steiger2021}, \cite{Nugent2023}, \cite{Cho2023}, and \cite{Chernozhukov2021} used aggregate indicators of mobility from mobile phone data records unattributed to specific travel modes to assess causality between mobility and COVID-19 cases. \cite{Steiger2021} estimated the causal impact of factors affecting the number of new cases of COVID-19 in Germany using a DAG and a negative binomial regression model. The authors found a positive causal impact between mobility in retail and recreational spaces and workplaces and COVID-19 cases, and a negative causal impact for essential shopping and mobility within residential areas. Other influential factors included weather, socio-demographic characteristics, and public awareness of COVID-19. \cite{Nugent2023} adopted the causal modified treatment policy method with the Super Learner machine learning algorithm to quantify the impact of mobility on COVID-19 cases in US counties. The authors found that unadjusted estimates suggested that decreasing mobility is causally linked to decreasing COVID-19 cases, but found that their adjusted estimates gave mixed results indicating potential confounding in the unadjusted estimates, which were most likely attributed to unknown differences in county characteristics. \cite{Cho2023} also analyzed the causal relationship between COVID-19 cases and mobility in the US. They adopted an ensemble empirical mode decomposition method with causal decomposition and reported that mobility levels have a causal relationship with long-term variations in COVID-19 cases rather short-term variations. \cite{Chernozhukov2021} used a causal structural model framework to test a range of counterfactual experiments including mobility changes and their impact on COVID-19 cases in the US. The authors find that reductions in mobility due to stay at home orders and business closures were effective in reducing case numbers, while the impacts of reductions in mobility due to school closures were inconclusive.

\subsection{Data}
The Community of Metros (COMET) benchmarking group, administered by the Transport Strategy Centre at Imperial College London, has provided ridership data from 8 international member metro networks. In keeping with commercial confidentiality requirements, the metro identities have been anonymized and are referred to with a numerical index ranging from 1 to 8 throughout this analysis. Therefore, the study units correspond to eight distinct geographical locations. The ridership data report the numerical ridership figures for each day. The data are recorded from before the onset of the COVID-19 pandemic to early 2022, with the earliest and latest dates in the dataset being 1 March 2020 and 23 January 2022. 

Data on COVID-19 case numbers correspond to the city that each metro is located in and refer to daily case numbers. The data on case numbers have been obtained from the Oxford University COVID-19 Government Response Tracker (OxCGRT) \citep{Hale2021}. Additional data from the OxCGRT database included in the analysis are the number of COVID-19 deaths per day, and a metric termed the Stringency Index which represents the severity of policy control measures implemented by governments during COVID-19 to limit transmission. The Stringency Index is a numeric index ranging from 0-100 with higher values corresponding to stricter control measures. We also include data on vaccinations which represent the cumulative total of vaccinations delivered on a daily basis from the Our World in Data database \citep{Mathieu2021}. Lastly, we include a series of variables representing all forms of mobility from Google Mobility \citep{Google2022}, specifically: residential, workplace, retail and recreation, grocery and pharmacy, and parks.

For context, a summary of all variables including their descriptive statistics aggregated across all metros in the analysis is given in Table \ref{tab:Descriptive statistics of variables}. The ridership is transformed in a logarithmic scale,  and the number of cases  are transformed with $\log(x+1)$ scale. From our preliminary analysis, the distribution of ridership is highly skewed and therefore we take the  $\log$ scale. After the transformation, the distribution looks more normal, and the ridership in $\log$ scale will be the treatment variable. For the outcome variable,  we will consider to use cases without transformation and in $\log(x+1)$ scale.

\begin{table}[ht]
	\centering
	\caption{Descriptive statistics of metro ridership data}
	\label{tab:Descriptive statistics of variables}
	\footnotesize
	\begin{tabular}{p{0.3\textwidth}p{0.3\textwidth}llll}
		\hline
		Variable & Description & Min. & Max. & Mean & Std. Dev. \\
		\hline
		Ridership ($log$  scale) & Number of passengers per day in log scale & 10.17  & 15.15 &13.02 & 1.07 \\
		Cases ($log(x+1)$  scale)& Number of confirmed COVID-19 cases in $log(x+1)$  scale & 4.52 & 14.31& 10.24 & 1.81 \\
		Deaths  & Number of confirmed COVID-19 deaths  & 12 & 121201 & 40167& 37874 \\
		Stringency  index & Stringency index ranging from 0-100 from OxCGRT. Represents degree of severity of policy measures to contain COVID-19 transmission. & 20.37 & 98.46 & 62.44 & 16.04 \\
		Total vaccinations per hundred & Number of total vaccinations i.e. sum of all doses per hundred people of population. For vaccines that require multiple doses, each individual dose is counted. & 0 & 238 & 56.40 & 71.87 \\
		Retail and recreation & Google mobility data for mobility changes relative to a baseline regular (pre-Covid) day for retail and recreation places & -97 & 36 & -37 & 22.33 \\
		Grocery and pharmacy & Google mobility data for mobility changes relative to a baseline regular (pre-Covid) day for grocery and pharmacy places & -91 & 99 & -9 & 20.16 \\
		Parks & Google mobility data for mobility changes relative to a baseline regular (pre-Covid) day for parks & -98 & 294 & -3 & 51.67 \\
		Transit stations & Google mobility data for mobility changes relative to a baseline regular (pre-Covid) day for transit stations & -93 & 61 & -41 & 21.80 \\
		Workplaces & Google mobility data for mobility changes relative to a baseline regular (pre-Covid) day for workplaces & -92 & 31 & -35 & 21.77 \\
		Residential & Google mobility data for mobility changes relative to a baseline regular (pre-Covid) day for residential places & -9 & 48 & 13 & 8.75 \\
			\hline
	\end{tabular}
\end{table}

\subsection{Results}
\begin{figure}[ht]
	\centering
	\includegraphics[width=1\linewidth]{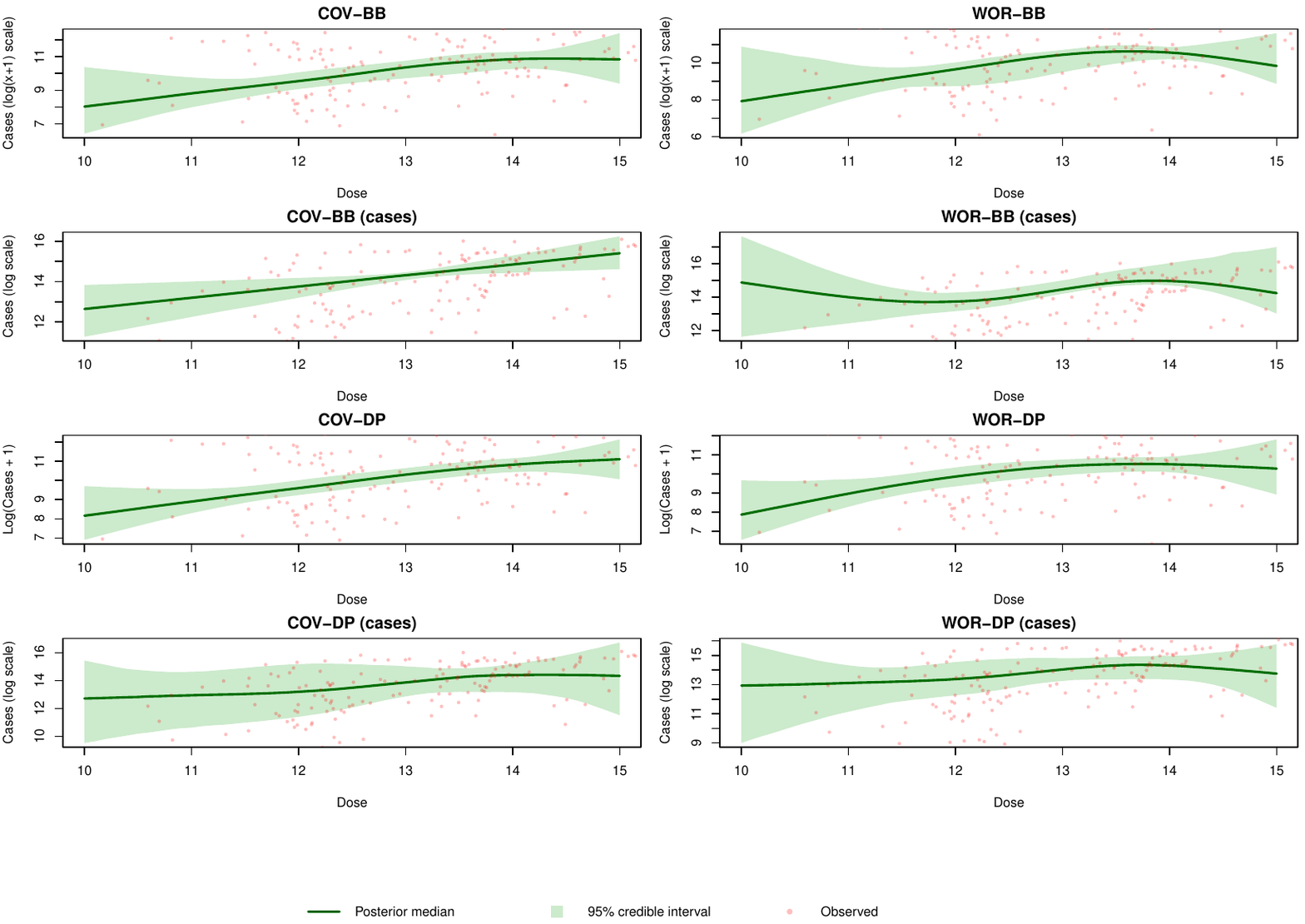}
	\caption{Application: Posterior predictive for dose–response estimates for ridership ($\log$ scale) and COVID-19 cases. Panels labeled $\log(x+1)$ use Gaussian GEE models for log-transformed cases, while panels labeled $\log$ scale use Poisson GEE models with posterior predictive samples rescaled to the $\log$ scale. Results are shown for Bayesian bootstrap  and DP inference under covariate-adjusted and weighted outcome regression specifications.}
	\label{real}
\end{figure}

To estimate the dose-response relationship between the ridership and COVID-19 cases, we employ the proposed Bayesian method to analyze two different outcomes: the raw case counts, modeled using a log link with the Poisson GEE, and the log-transformed case counts, $\log(x+1)$, modeled under an identity link with the Gaussian GEE. For both outcomes, we specify an independent working correlation structure and included all relevant confounders along with a cubic spline function of the estimated GPS as covariates. Similar to the simulation, for the GPS model, we use a GEE model that adjusts for all time-varying confounders. We generate a total of 5,000 posterior samples. For the DP method, we generate a synthetic dataset of size $N=200$ using a procedure similar to that employed in the simulation studies. Due to the relatively small sample size, the stick-breaking weights in the DP decrease rapidly, falling below $10^{-8}$ after approximately 200 iterations. Figure \ref{real} presents posterior predictive dose-response curves estimated using COV and WOR methods.  Across all specifications, the estimated dose-response relationship indicates an overall positive and monotone causal effect of mobility on COVID-19 cases, with higher ridership associated with higher expected case counts. Under the COV-based approach, the estimated dose-response function increases for both log-transformed and count outcomes, although uncertainty is larger in certain dose ranges (notably between 10–12 and 14–15) due to reduced data density. The DP-based approach shows a more stable curve as the synthetic dataset has a larger sample. When using case counts as the outcome, the COV-based method shows a generally exponential increase in cases as the logarithm of ridership rises. However, the DP-based methods appear to overcompensate in certain regions, resulting in some non-smooth curves likely because some portion of the resampled data is drawn from areas that deviate from the main trend. In contrast, the WOR estimator shows a rapid increase in cases at lower dose levels followed by a plateau and slight decline at higher doses, likely due to sensitivity to reweighting in sparsely populated regions.  In regions with sparse data, the DP-based approach produces more stable estimates, suggesting improved performance in low-density areas. In summary, case counts tend to increase with the dose level, which suggests that public transport usage rises, but this relationship is not strictly linear. While both the log-transformed and count-based outcomes suggest a strong upward trend in case numbers at lower dose regions, the log-transformed model shows a more moderate increase because of its smoothing effect, whereas the count model captures a steeper, non-smooth and exponential rise.

\section{Discussion}
\label{sec:dis}
In this paper, we introduce a non-parametric Bayesian framework for estimating longitudinal causal dose-response relationships in the presence of repeated measures by adopting a GEE perspective. Instead of specifying a fully parametric likelihood, the target estimand is identified via parameters defined in moment conditions, and uncertainty is quantified by placing a nonparametric prior on the underlying data-generating distribution. Specifically, this approach extends traditional Bayesian methodology by integrating the generalized Bayesian bootstrap within a GEE framework, thereby explicitly accounting for the temporal structure of longitudinal data. Both the GPS and the outcome mean are modeled using GEEs, and the GPS is incorporated into the outcome model in two ways: through  covariate adjustment or stabilized inverse weighting. Simulation results indicate that both Bayesian dose-response estimators are approximately unbiased and achieve near-nominal posterior coverage across a range of scenarios when using the DP specification. However, the weighting-based estimator exhibits larger posterior variance, a behavior consistent with other inverse-weighting approaches for continuous treatments, where exposure binning is often required to mitigate variability \citep{naimi2014constructing}. In our empirical study, we investigate the impact of metro ridership on COVID-19 case counts across major urban centers during the pandemic.  The results suggest that increased metro ridership contributes to the spread of COVID-19 up to a certain exposure threshold, beyond which the effect plateaus. These findings have important implications for pandemic-related urban policy, particularly in informing lockdown strategies and optimizing transportation systems in future public health crises.

One appealing aspect of Bayesian estimators is that they permit flexible modelling with minimal assumptions about the dose-response relationship and allow prior knowledge to be incorporated into the model. The proposed method can also easily handle conditionally censored data by incorporating time-varying censoring weights.  Bayesian methods are increasingly influential in applied causal inference, offering the ability to make direct probabilistic statements about treatment effects and to assess sensitivity with respect to alternative priors or expert-driven inputs. Despite these advantages, Bayesian causal methods for longitudinal data remain underdeveloped. In this paper, we proposed a scalable Bayesian approach that effectively handles longitudinal data and is easily implementable.  In our framework, we assume there is no unmeasured confounding beyond the observed pre-exposure covariates $X_t$. We use $U$ to represent unobserved time-invariant heterogeneity in the outcome process, and we assume that dose assignment is fully explained by $X_t$. If instead there are latent factors that jointly influence both treatment and outcomes beyond what is captured by $X_t$, then the GPS based only on observed covariates will be insufficient for identification. Addressing such settings would likely require joint modeling approaches with shared latent structures or correlated random effects \citep{shardell2018joint}. Therefore, extending the proposed Bayesian longitudinal dose-response framework to accommodate latent confounding remains an important direction for future research.  In addition, our proposed Bayesian dose-response estimation is derived under the SUTVA assumption, assuming no interference. In our motivating COVID study, this assumption translates to each unit’s monthly metro-ridership exposure being assumed to affect only its own subsequent COVID-19 risk, which can be violated in application. Methodological extensions that relax SUTVA are emerging \citep{o2025local}. Future work will adapt these ideas to longitudinal, continuous-dose settings, enabling joint estimation of direct and indirect (spillover) ridership effects on COVID-19 dynamics. Despite these limitations, the proposed estimator fills an important methodological gap by coupling non-parametric Bayesian inference with repeated-measures data and a time-varying continuous exposure. It remains fully appropriate for applications where interference is unlikely (e.g., individual risk-factor studies in chronic-disease epidemiology) and rich observed confounder information is available, such as many longitudinal epidemiologic and public health studies. Furthermore, our framework offers a robust foundation that broader spillover-aware extensions can build upon.

\section*{Acknowledgement}
Yu Luo was supported by the Engineering \& Physical Sciences Research Council (EPSRC) Grant EP/Y029755/1. Kuan Liu was supported by the Canadian Institutes of Health Research (CIHR) Grant AD7200183.

\appendix
\section{Unbiasedness of the GPS-Weighted GEE}
\label{app:canonical}
The working covariance matrix in a GEE framework is typically defined as
\begin{equation*}
	\Sigma_i = \mathbf{A}_i^{1/2} \mathbf{R}_i(\alpha) \mathbf{A}_i^{1/2},
\end{equation*}
where $\mathbf{A}_i = \text{diag}\{v(\nu_{it})\}$ is a diagonal matrix of variance functions, and $\mathbf{R}_i(\alpha)$ is the working correlation matrix. Under the independence working correlation assumption, $\mathbf{R}_i(\alpha) = \mathbf{I}_{T_i\times T_i}$. This simplifies the inverse covariance to a diagonal matrix, i.e.,
\begin{equation*}
	\Sigma_i^{-1} = \mathbf{A}_i^{-1} = \text{diag} \left( \frac{1}{ v(\nu_{i1})}, \dots, \frac{1}{ v(\nu_{iT_i})} \right).
\end{equation*}
Substituting the diagonal matrix $\Sigma_i^{-1}$ and the diagonal weight matrix $\mathbf{w}_i = \text{diag}(w_{i1}, \dots, w_{iT_i})$ into the GEE, we obtain
\begin{equation*}
	\mathbf{U}_i^W(\xi) = \left( \frac{\partial \nu^{W}_i(\xi)}{\partial \xi} \right)^\top \text{diag} \left( \frac{1}{ v(\nu_{it})} \right) \text{diag}(w_{it}) \left( Y_i - \nu^{W}_i(\xi) \right).
\end{equation*}
Because all intermediate matrices are diagonal, the $t$-th element of the derivative and weight only interact with the $t$-th element of the residual vector. Thus, the matrix product collapses into a summation over time, that is, 
\begin{equation*}
	\mathbf{U}_i^W(\xi) = \sum_{t=1}^{T_i} \left( \frac{\partial \nu^{W}_{it}(\xi)}{\partial \xi} \right)^\top \frac{w_{it}}{ v(\nu_{it})} \left( Y_{it} - \nu^{W}_{it}(\xi) \right).
\end{equation*}
Let $\mathbf{H}_{it} = \left( \frac{\partial \nu^{W}_{it}(\xi)}{\partial \xi} \right)^\top \frac{1}{ v(\nu_{it})}$. Note that $\mathbf{H}_{it}$ is a function of the treatment $D_{it}$ and parameters $\xi$, and is independent of the random variation in $Y_{it}$. Conditioning on $X_{it}$ and $D_{it}$, we have
\begin{equation*}
	\E[\mathbf{H}_{it} w_{it} (Y_{it} - \nu^{W}_{it}(\xi))] = \E_{X,D} \left[ \mathbf{H}_{it} w_{it} \E[Y_{it} - \nu^{W}_{it}(\xi) \mid X_{it}, D_{it}] \right].
\end{equation*}
Substituting the GPS weight $w_{it} = \frac{f(D_{it})}{f(D_{it} \mid X_{it})}$ and expressing the outer expectation as an integral
\begin{equation*}
	\int \int \mathbf{H}_{it} \frac{f(D_{it})}{f(D_{it} \mid X_{it})} \left( \E[Y_{it} \mid X_{it}, D_{it}] - \nu^{W}_{it}(\xi) \right) f(D_{it} \mid X_{it}) f(X_{it}) \, dD_{it} dX_{it}.
\end{equation*}
The conditional density $f(D_{it} \mid X_{it})$ cancels out, yielding
\begin{equation*}
	\int \mathbf{H}_{it} f(D_{it}) \underbrace{\left[ \int (\E[Y_{it} \mid X_{it}, D_{it}] - \nu^{W}_{it}(\xi)) f(X_{it}) \, dX_{it}\right]}_{\text{Term A}} \, dD_{it}.
\end{equation*}
By the identification  of the marginal APO
\begin{equation*}
	\int \E[Y_{it} \mid X_{it}, D_{it}=d] f(X_{it}) \, dX_{it} = \mu(d).
\end{equation*}
Since $\nu^{W}_{it}(\xi)$ is the model for $\mu(d)$, Term A equals 0 at the true parameter value, i.e., $\E[\mathbf{U}_i^W(\xi_0)] = 0$.  

If $\mathbf{R}_i(\alpha)$ were non-diagonal, $\Sigma_i^{-1}$ would involve off-diagonal elements where a weight $w_{is}$ from time $s$ would multiply a residual from time $t \neq s$. Because $w_{is}$ is not designed to unconfound covariates at time $t$, such terms would not generally vanish in expectation.

\section{Interventional definition and observational identification of $\mu(d)$}
\label{app:iden}
The causal estimand is $\mu_t(d)=\mathbb{E}_{\mathcal{E}_d}(Y_t)$, and under SUTVA and consistency $\mathbb{E}_{\mathcal{E}_d}(Y_t)=\mathbb{E}\{Y_t(d)\}$.
Under ignorability and positivity (Section~\ref{sec:apo}), $\mu_t(d)$ is identified from $\mathcal{O}$ via the GPS representation:
\[
\mu_t(d)=\mathbb{E}_{X_t}\!\left[
\mathbb{E}\!\left\{Y_t \mid D_t=d,\ e_t(d,X_t)\right\}
\right],
\qquad d\in\mathcal D.
\]
If the working MSM $g\{\mu(d)\}=m(d;\xi)$ is correctly specified and $\mathbf U^{W}(\xi)$ denotes the corresponding GEE estimating function constructed from the identified observed-data representation in Section~\ref{sec:apo}, then $\mathbf U^{W}(\xi)$ is unbiased at the causal MSM parameter $\xi^\star$, i.e.,
\[
\mathbb{E}_{\mathcal{O}}\{\mathbf U^{W}(\xi^\star)\}=0.
\]
The population moment restriction in Section~\ref{Sec:bb} is an expectation under $F_0\equiv \mathcal{O}$ while retaining a causal interpretation for $\psi=\xi$.

\bibliographystyle{chicago}
\bibliography{reference}
\end{document}